\documentclass[preprint]{article}
\usepackage{spconf,amsmath,graphicx}
\usepackage{amsfonts}
\usepackage{booktabs}
\usepackage[colorlinks=true]{hyperref}
\usepackage{multirow}
\usepackage{subcaption}
\usepackage{tikz}
\usepackage{amsmath,graphicx}
\usepackage{algpseudocode}
\usepackage{algorithm}
\usepackage{color}
\usepackage{tabularray}
\usepackage{booktabs}
\usepackage{adjustbox}
\usepackage{colortbl}

\copyrightnotice{979-8-3503-0689-7/23/\$31.00~\copyright2023 IEEE}

\title{VoiceExtender: Short-utterance Text-independent Speaker Verification with Guided Diffusion Model}
\name{Yayun He$^*$, Zuheng Kang$^*$\thanks{$^*$Equal contributions}, Jianzong Wang$^\dagger$\thanks{ $^\dagger$Corresponding author: Jianzong Wang, jzwang@188.com}, Junqing Peng, Jing Xiao}
\address{Ping An Technology (Shenzhen) Co., Ltd., Shenzhen, China}
\begin{document}
%
\maketitle
\begin{abstract}

  Speaker verification (SV) performance deteriorates as utterances become shorter.
  To this end, we propose a new architecture called VoiceExtender which provides a promising solution for improving SV performance when handling short-duration speech signals.
  We use two guided diffusion models, the built-in and the external speaker embedding (SE) guided diffusion model, both of which utilize a diffusion model-based sample generator that leverages SE guidance to augment the speech features based on a short utterance.
  Extensive experimental results on the VoxCeleb1 dataset show that our method outperforms the baseline, with relative improvements in equal error rate (EER) of 46.1\%, 35.7\%, 10.4\%, and 5.7\% for the short utterance conditions of 0.5, 1.0, 1.5, and 2.0 seconds, respectively.

\end{abstract}

\begin{figure*}[t]
  \includegraphics[width=0.95\textwidth]{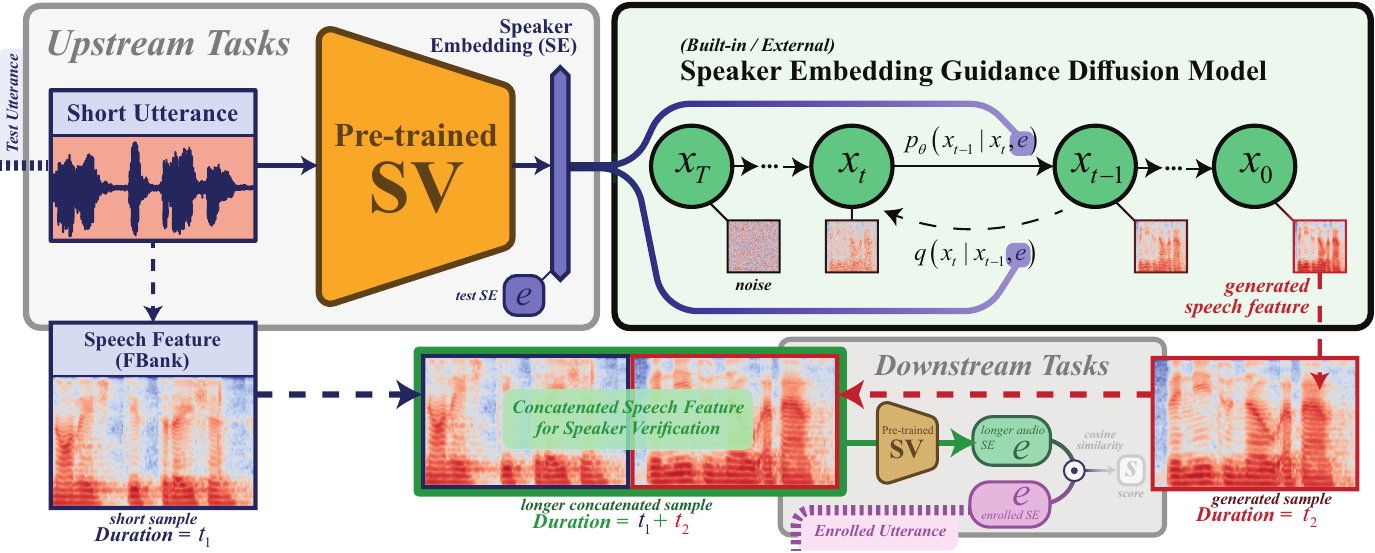}
  \centering
  \caption{VoiceExtender framework overview: speaker embedding guidance diffusion model.}
  \label{fig:overview}
\end{figure*}

\section{Introduction}
\label{sec:intro}

Speaker verification (SV) plays a vital role in several key areas such as security authentication, privacy protection, and live detection, where it can provide reliable evidence or personalized services.
Although the current state-of-the-art SV models \cite{Reynolds2000SpeakerVU,snyder2018x,desplanques2020ecapa,Wang2023CAMAF} perform excellent with long utterances, they are far from satisfactory on short utterances.
Especially in certain situations, such as criminal investigation and fraud detection, it is often difficult to ensure that a long enough voice can be obtained from the speaker.
However, the performance of SV can be terrible for very short utterances.
Extensive experimental results have shown that there exists a certain limit to the utterance duration, which is usually 2 seconds: when the utterance duration is less than this limit, the performance of SV is severely degraded \cite{poddar2018speaker,Jung2018ShortUC}.
Therefore, it is an important and open research problem to explore how to enhance the SV performance under the conditions of very short utterance duration.

In recent years, many researchers have attempted to address the problem of short-utterance SV.
Some researchers use more sophisticated neural network structures to improve performance during training.
A. Hajavi et al. \cite{hajavi2019deep} propose a novel architecture that is more suitable for the short-utterance SV problem by mixing convolutional neural networks, and residual networks, with complex connectivity.
J. King et al. \cite{kim2022rawnext} perform iterative and hierarchical aggregation of features of different lengths and apply dynamic block scaling to accommodate different audio durations in SV.
J.w. Jung et al. \cite{jung2019short} employ a teacher-student framework that compensates for the use of additional phonetic-level and utterance-level features throughout the network.
W. Chen et al. \cite{chen2020length} improve the SV performance on short utterances by enhancing x-vector \cite{snyder2018x} with more advanced self-attention, pooling structures, and more sophisticated loss techniques.
Some researchers achieve even better results by using more advanced self-supervised learning features instead of artificial ones, such as Mel-filterbanks (FBanks).
W.H. Sang et al. \cite{Han2023ShortSegmentSV} have achieved great results on shorter utterances using novel features generated by a multi-resolution encoder.
Other researchers have used speaker embedding-based meta-learning methods to incorporate native pre-trained SV models with powerful short-utterance SV capabilities.
K. Liu et al. \cite{liu2020text} propose an embedding mapping model with Wasserstein generative adversarial network (W-GAN) \cite{arjovsky2017wasserstein}, which yields enhanced embeddings from short utterances and improves recognizability.
N. Tawara et al. \cite{tawara2020frame} introduce an adversarial framework to obtain phoneme-invariant speaker embeddings at the frame level in order to overcome performance degradation.
In summary, these methods, no matter how they are applied, all consider how to better exploit and mine the useful information contained in a given utterance.
However, general SV frameworks perform poorly on short utterances, not only because the information in the short utterance is underutilized, but also because the utterances themselves are too short and do not contain enough information.
If it is possible to make these utterances longer using audio generation methods so that more effective information becomes available, then these generated audios should be able to help improve SV performance.

Today, artificial intelligence generated content (AIGC) has profoundly reshaped the creation and consumption of media, art, education, and entertainment.
Advanced AIGC models are capable of generating high-quality, human-like text, image, audio, and video, with the potential to augment human creativity and improve the efficiency of content production.
It not only allows people to focus on more meaningful things but also greatly expands access to personalized content.
Since the AIGC method is capable of producing high-quality and reliable content under given conditions, it is also capable of generating speech with given speaker identity information, thus improving SV performance.
Popular and powerful AIGC methods such as VqVAE \cite{yayun2023dance}, GAN \cite{Brock2018LargeSG,Kong2020HiFiGANGA}, and flow-based methods (e.g., GLOW \cite{Prenger2018WaveglowAF,Kingma2018GlowGF}) have been successfully applied to image, audio, and video generation.
Transformer-based methods have been successfully applied to sequence generation, such as text \cite{Aljanabi2023ChatGPTFD}, music \cite{Huang2021PopPM,mittal2021symbolic}, and dance generation \cite{Siyao2022Bailando3D,yayun2023dance}.
If such a powerful generative capability could be used to generate speech, this additional information would help SV.

The diffusion model achieves state-of-the-art sample quality on image \cite{rombach2022high,lugmayr2022repaint,ruiz2022dreambooth,rombach2022text,xu2022versatile}, speech or singing \cite{Liu2021DiffSingerSV}, and video \cite{ho2022video,Zhang2023ControlVideoTC} synthesis benchmarks.
Compared to other methods, such as generative adversarial networks (GAN) \cite{goodfellow2020generative,creswell2018generative}, which generate poor sample diversity and are prone to pattern collapse, flow- and VAE-based models, which generate lower quality and insufficiently controllable samples, the diffusion model generates samples with more diversity and controllability, and the training process is more stable.
To generate controllable data, a guided diffusion model for data synthesis under given conditions is proposed.
P. Dhariwal et al. \cite{dhariwal2021diffusion} present the classifier guidance approach, which uses externally trained classifiers to generate images of specified categories, and effectively improves the quality of generated samples.
J. Ho et al. \cite{ho2022classifier} introduce the classifier-free guidance technique, which mixes scores estimated from conditional and unconditional diffusion models, and proves that the guidance diffusion model can be performed nicely without the classifier.

Speaker embeddings (SE) are most effective in distinguishing the identity of the speaker.
Inspired by the guidance diffusion model, we adopt SE as guidance for high-quality speech synthesis.
Then, the final utterance can become longer by concatenating the original utterance and the generated utterance for better SV.
This voice generation and concatenation process behaves like a voice extension process, so we name this framework a VoiceExtender.

Based on the above analysis, we make the following contributions:
(1) We present VoiceExtender, a promising framework for improving the performance of short-utterance speaker verification (SV) by extending the utterance length using diffusion models;
(2) We employ a speaker embedding (SE) guided diffusion model technique, which enables the generation of speech samples with deep features of the specified speaker;
(3) Extensive experiments have shown the effectiveness of VoiceExtender on VoxCeleb, especially in extremely short utterance scenarios.

\section{Methodology}
\label{sec:method}

\subsection{Speaker Embedding Guidance Diffusion Model}
\label{sec:overview}

Recent advances in diffusion models \cite{yang2022diffusion,cao2022survey,song2020denoising,nichol2021improved,ho2022video} have demonstrated their ability to generate high-quality images while capturing the full data distribution.
In the speaker verification (SV) task, diffusion models can similarly learn to model the overall distribution of a speaker's utterances.
By conditioning the diffusion process on speaker embeddings (SEs) extracted from a reference utterance, the model could plausibly generate new synthetic speech that captures the vocal characteristics of that speaker.
In other words, the strong generative capabilities of diffusion models, when guided by SEs during sampling, may enable the synthesis of new speech segments that closely match the speaker identity of a given reference utterance.
This opens up new opportunities for speaker-conditional speech synthesis using diffusion models for SV.

Gaussian diffusion models include two processes:
(1) the forward process from data to Gaussian noise and
(2) the backward process from noise to data.
In the forward process, the noisy data $ x_t $ at each step $ t $ can be obtained by adding Gaussian noise to the data at step $ t-1 $ as shown in Equation \ref{eq:gaussian_forward}.
Where $ 1-\alpha _t $ is the magnitude of the noise needed to be added at each step.

\noindent
\begin{equation}
  \label{eq:gaussian_forward}
  q\left( x_t\left| x_{t-1} \right. \right) =\mathcal{N}\left( x_t;\sqrt{\alpha _t}x_{t-1},\left( 1-\alpha _t \right) \mathcal{I} \right)
\end{equation}
\noindent

In the reverse process, the posterior $
  q\left( x_{t-1}\left| x_t \right. \right)
$ can be approximated by a model $
  p_{\theta}\left( x_{t-1}\left| x_t \right. \right)
$ as shown in Equation \ref{eq:gaussian_reversed}.
Where $ \mu _{\theta}\left( x_t \right) $ is the mean value of the sample at step $ t-1 $ calculated from $ x_{t} $.
$ \varSigma _{\theta}\left( x_t \right) $ is a value related to $ t $ and $ \alpha _t $ and it's value has nothing to do with $ x_t $ actually.

\noindent
\begin{equation}
  \label{eq:gaussian_reversed}
  p_{\theta}\left( x_{t-1}\left| x_t \right. \right) =\mathcal{N}\left( \mu _{\theta}\left( x_t \right) ,\varSigma _{\theta}\left( x_t \right) \right)
\end{equation}
\noindent

The synthesized data $ x_0 $ can be obtained step by step by starting with Gaussian noise $ x_T $.
In order to get $ \mu _{\theta}\left( x_t \right) $ and $ \varSigma _{\theta}\left( x_t \right) $, J. Ho et al. \cite{ho2020denoising} show that we can derive $ \mu _{\theta}\left( x_t \right) $ from a noise estimation model $ \epsilon _{\theta}\left( x_t,t \right) $ which can be trained during the process of adding noise with a standard mean-squared error loss $ \mathcal{L}_{\mathrm{simple}} $ defined in Equation \ref{eq:loss_train}, where $ \epsilon $ is the Gaussian noise added from $ x_0 $ to $ x_t $.

\noindent
\begin{equation}
  \label{eq:loss_train}
  \mathcal{L}_{\mathrm{simple}}=\mathbb{E}_{x_0,t,\epsilon}\left[ \lVert \epsilon -\epsilon _{\theta}\left( x_t,t \right) \rVert ^2 \right]
\end{equation}
\noindent

The relationship between $ \mu _{\theta}\left( x_t \right) $ and $
  \epsilon _{\theta}\left( x_t,t \right) $ is shown in Equation \ref{eq:relation}.
Where $ \bar{\alpha}_t $ is the accumulated multiplication of $ \left( \alpha _1,\alpha _2,\alpha _3...\alpha _t \right) $.

\noindent
\begin{equation}
  \label{eq:relation}
  \mu _{\theta}\left( x_t \right) =\frac{1}{\sqrt{\alpha _t}}\left( x_t-\frac{1-\alpha _t}{\sqrt{1-\bar{\alpha}_t}}\epsilon _{\theta}\left( x_t,t \right) \right)
\end{equation}
\noindent

Previous studies have demonstrated that speaker verification (SV) models exhibit significant performance degradation when evaluated on exceedingly short test utterances.
To address this challenge, we propose a VoiceExtender framework (illustrated in Figure \ref{fig:overview}) that leverages a guided diffusion model to synthesize augmented speech features enriched with supplementary speaker information.
The generated speech features are then integrated into the original test utterance, effectively extending its duration and improving the verification quality.
To achieve this, we present and compare two implementations of VoiceExtender that differ in how speaker embeddings (SE) are applied during sample generation:
(1) An external SE diffusion model (discussed in $\S$ \ref{sec:EDM}), where SE is provided as the conditional input to the diffusion model.
(2) A built-in SE diffusion model (detailed in $\S$ \ref{sec:BDM}), where SE is fused into the diffusion model.

\subsection{External Speaker Embedding Diffusion Model}
\label{sec:EDM}

\begin{figure}[t]
  \includegraphics[width=0.49\textwidth]{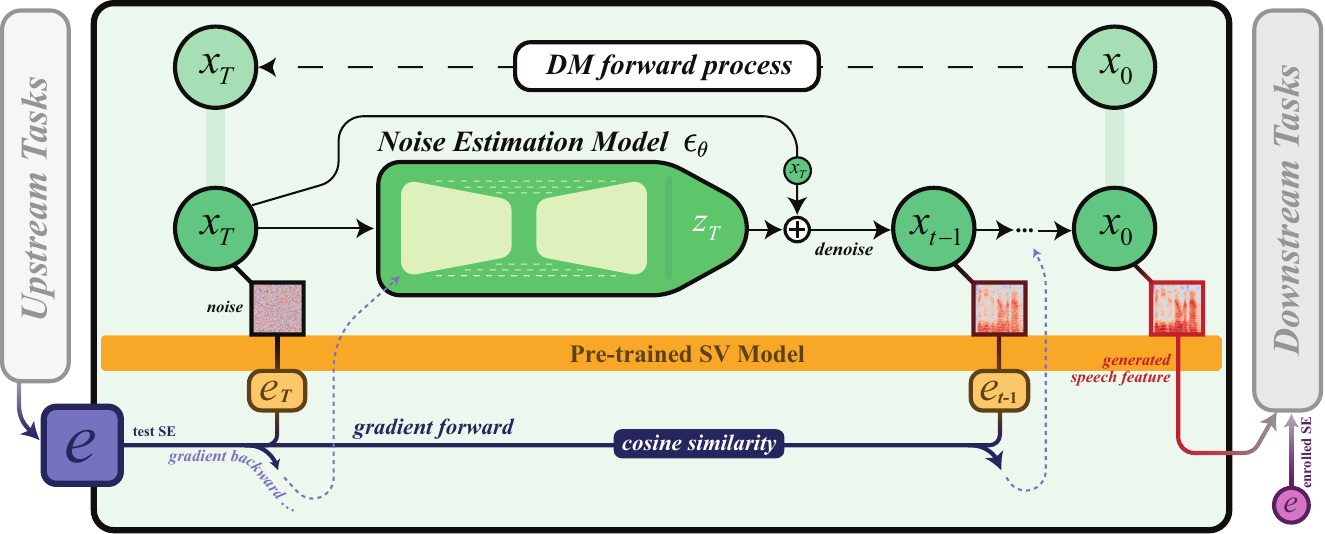}
  \centering
  \caption{External SE diffusion model.}
  \label{fig:external}
\end{figure}

P. Dhariwal et al. \cite{dhariwal2021diffusion} demonstrate diffusion models conditioned on classifier guidance to generate class-specific images.
Conditioning is achieved by influencing the diffusion model's mean prediction $\mu_\theta(x_t | y)$ with the gradient of the classifier's predicted log-probability $\log p_\phi(y|x_t)$ for class $y$.
Inspired by this, we propose a method to generate new samples using a diffusion model guided by the similarity between SEs.
Specifically, given a short reference speech input, we aim to synthesize a new speech feature with an SE close to the reference.
Further details are shown in Figure \ref{fig:external}.

As illustrated in the figure, our proposed framework first leverages a pre-trained SV model to extract a SE $ e $ from a given reference utterance.
This SE is then utilized to guide the sampling process of an external SE-conditioned diffusion model.
Since the SEs are normalized, the similarity between the reference SE $ e $ and the SE $ f_\phi(x_t) $ predicted by the model for a partially diffused sample $ x_t $ can be efficiently computed as their dot product $ f_\phi(x_t) \cdot e $.
A larger dot product indicates a greater similarity between the two embeddings.
The gradient of this dot product with respect to $x_t$, i.e. $ \nabla_{x_t} (f_\phi(x_t) \cdot e) $, is then utilized to guide the next sampling step, steering the model to generate samples with SEs closer to the reference.
In this way, the external SE guidance can control the characteristics of the synthesized speech.

The reverse diffusion process can be conditioned on a reference SE $ e $ by incorporating a similarity loss between $ e $ and the embeddings predicted for generated samples.
Then, the perturbed reverse process mean $ \hat{\mu}_{\theta}\left( x_t\left| e \right. \right) $ is defined in Equation \ref{eq:reverse_mean}, where $ s $ is the gradient scale.
$ x_{t-1} $ can be sampled from
$ \mathcal{N}\left( \mu +s\varSigma \nabla _{x_t}\left( f_{\phi}\left( x_t \right) \cdot e \right) ,\varSigma \right) $.

\noindent
\begin{equation}
  \label{eq:reverse_mean}
  \hat{\mu}_{\theta}\left( x_t\left| e \right. \right) =\mu _{\theta}\left( x_t \right) +s\cdot \varSigma _{\theta}\left( x_t \right) \nabla _{x_t}\left( f_{\phi}\left( x_t \right) \cdot e \right)
\end{equation}
\noindent

\begin{algorithm}[ht]
  \floatname{algorithm}{Algorithm}
  \renewcommand{\algorithmicrequire}{\textbf{Input:}}
  \renewcommand{\algorithmicensure}{\textbf{Output:}}
  \caption{External SE diffusion model sampling}
  \label{pseudo:fic}
  \footnotesize
  \begin{algorithmic}[1]
    \Require guidance SE $ e $, gradient scale $ s $
    \Ensure $ x_0 $

    \State Given: diffusion model $
      \left( \mu _{\theta}\left( x_t \right) ,\varSigma _{\theta}\left( x_t \right) \right)
    $, SE model $
      f_{\phi}\left( x \right)
    $
    \State $ x_T\gets $ sample from $ \mathcal{N}\left( 0,I \right) $
    \For{$ t=\ T,\cdots 1 $}
    \State $ \mu ,\varSigma \gets \mu _{\theta},\sigma _{\theta}^2I $
    \State $
      x_{t-1}\gets $ sample from $\mathcal{N}\left( \mu +s\varSigma \nabla _{x_t}\left( f_{\phi}\left( x_t \right) \cdot e \right) ,\varSigma \right)
    $
    \EndFor
    \State \Return $ x_0 $
  \end{algorithmic}
\end{algorithm}

The sampling procedure for the proposed external SE diffusion model is detailed in Algorithm \ref{pseudo:fic}.
At each reverse diffusion timestep $ t $, the pre-trained SV model first extracts the SE $ e_t $ from the currently generated sample $ x_t $.
The similarity between $ e_t $ and the reference SE $ e $ is then computed to ensure that the model generation process is correct.

\subsection{Built-in Speaker Embedding Diffusion Model}
\label{sec:BDM}

J. Ho et al. \cite{ho2022classifier} have proposed a classifier-free guidance approach for conditioned diffusion models.
Rather than relying on an external classifier, this method directly embeds the class label $ y $ into the model's noise estimation $ \epsilon_\theta(x_t|y) $.
With this structure, the trained model can support both class-conditional and unconditional sampling.
During sampling, the noise prediction is a combination of conditional and unconditional estimation, as shown in Equation \ref{eq:classifier_free_guide}.
The gradient can be written in Equation \ref{eq:gradient}.

\noindent
\begin{equation}
  \label{eq:classifier_free_guide}
  \hat{\epsilon}_{\theta}\left( x_t\left| y \right. \right) =\epsilon _{\theta}\left( x_t \right) +s\cdot \left( \epsilon _{\theta}\left( x_t\left| y \right. \right) -\epsilon _{\theta}\left( x_t \right) \right)
\end{equation}
\noindent

\noindent
\begin{equation}
  \label{eq:gradient}
  \begin{split}
    \nabla _{x_t}\log p^i\left( x_t\left| y \right. \right) &\propto \nabla _{x_t}\log p\left( x_t\left| y \right. \right) -\nabla _{x_t}\log p\left( x_t \right) \\
    &\propto \epsilon ^*\left( x_t\left| y \right. \right) -\epsilon ^*\left( x_t \right)
  \end{split}
\end{equation}

To implement the built-in SE diffusion model, the class label $ y $ in Equation \ref{eq:classifier_free_guide} can be replaced by the guide information of SE $ e $.
The prediction $ \hat{\epsilon}_{} $ is defined in Equation \ref{eq:e_free_guide}.
The $ \mu _{\theta}\left( x_t \right) $ can be obtained according to Equation \ref{eq:relation}.
The detailed sampling procedure for the built-in SE diffusion model is shown in the pseudocode of the Algorithm \ref{pseudo:fic2}.

\noindent
\begin{equation}
  \label{eq:e_free_guide}
  \hat{\epsilon}_{\theta}\left( x_t\left| e \right. \right) =\epsilon _{\theta}\left( x_t \right) +s\cdot \left( \epsilon _{\theta}\left( x_t\left| e \right. \right) -\epsilon _{\theta}\left( x_t \right) \right)
\end{equation}
\noindent

\begin{algorithm}[ht]
  \floatname{algorithm}{Algorithm}
  \renewcommand{\algorithmicrequire}{\textbf{Input:}}
  \renewcommand{\algorithmicensure}{\textbf{Output:}}
  \caption{Built-in SE diffusion model sampling}
  \label{pseudo:fic2}
  \footnotesize
  \begin{algorithmic}[1]
    \Require guidance SE $ e $, gradient scale $ s $
    \Ensure $ x_0 $

    \State Given: diffusion model without e $
      \epsilon _{\theta}\left( x_t \right)
    $, diffusion model with e $
      \epsilon _{\theta}\left( x_t\left| e \right. \right)
    $
    \State $ x_T\gets $ sample from $ \mathcal{N}\left( 0,I \right) $
    \For{$ t=\ T,\cdots 1 $}
    \State $ \hat{\epsilon}_{\theta}\left( x_t\left| e \right. \right) =\epsilon _{\theta}\left( x_t \right) +s\cdot \left( \epsilon _{\theta}\left( x_t\left| e \right. \right) -\epsilon _{\theta}\left( x_t \right) \right) $
    \State $ \mu =\frac{1}{\sqrt{\alpha _t}}\left( x_t-\frac{1-\alpha _t}{\sqrt{1-\bar{\alpha}_t}}\hat{\epsilon}_{\theta}\left( x_t|e \right) \right) $
    \State $
      x_{t-1}\gets $ sample from $\mathcal{N}\left( \mu,\varSigma \right)
    $
    \EndFor
    \State \Return $ x_0 $
  \end{algorithmic}
\end{algorithm}

\begin{figure}[t]
  \includegraphics[width=0.49\textwidth]{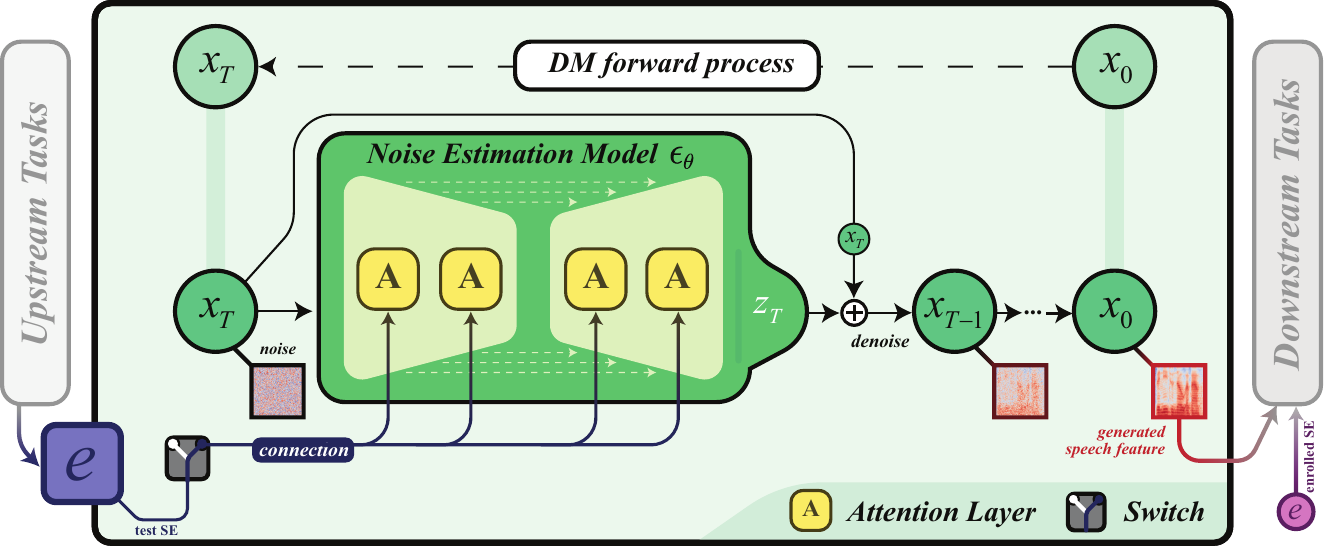}
  \centering
  \caption{Built-in SE diffusion model.}
  \label{fig:buildin}
\end{figure}

Figure \ref{fig:buildin} shows the process of the built-in SE diffusion model.
As can be seen from the figure, we borrowed the architecture of the ablated diffusion model (ADM) proposed by \cite{dhariwal2021diffusion} and extended it with SE as a condition.
Since the base model of ADM uses a UNet with an attention module \cite{vaswani2017attention}, it performs nicely in synthesis tasks.
The ADM in our work performs noise estimation in the inverse process.
At each attention layer of the ADM model, the guidance SE $ e $ is first mapped to an appropriate dimension and then concatenated to the context of these attention layers.

\section{Experiments}
\label{sec:experiment}

\subsection{Dataset}
\label{sec:dataset}

The training and testing of the diffusion model are accomplished based on VoxCeleb1 \cite{nagrani2017voxceleb}.
The training set contains 1,100 speakers with 134,941 utterances, and the test set has 151 speakers and 18,575 utterances.
There are 50,000 test pairs used for the evaluation.
In all tests in the experiment, the test pairs of utterances are the same for each test, only the duration of the utterances differs.
The ratio of test pairs for the same speaker and different speakers is 1:1.
More details can be found in Table \ref{tab:vox}.

\begin{table}[ht]
  \centering
  \footnotesize
  \renewcommand{\arraystretch}{1.0}
  \setlength{\tabcolsep}{26.2pt}
  \begin{tabular}{@{}lcc@{}}
    \toprule
    \textbf{Data}  & \textbf{Num of speakers} & \textbf{Num of utterances} \\ \midrule
    \textbf{Train} & 1,100                    & 134,941                    \\
    \textbf{Test}  & 151                      & 18,575                     \\
    \textbf{Total} & 1,251                    & 153,516                    \\ \bottomrule
  \end{tabular}
  \caption{Details of the VoxCeleb1 dataset.}
  \label{tab:vox}
\end{table}

\begin{table*}[t]
  \centering
  \footnotesize
  \setlength{\tabcolsep}{9.9pt}
  \renewcommand{\arraystretch}{1.0}
  \begin{adjustbox}{max height=0.7\textheight}
    \begin{tabular}{@{}llcccccccc@{}}
      \toprule
      \multirow{2}{*}{\textbf{Model Type}}                                                                     &                        & \multicolumn{2}{c}{\textbf{0.5s}} & \multicolumn{2}{c}{\textbf{1.0s}} & \multicolumn{2}{c}{\textbf{1.5s}} & \multicolumn{2}{c}{\textbf{2.0s}}                                                                           \\ \cmidrule(l){2-10}
                                                                                                               &                        & \textbf{EER(\%)}                  & \textbf{MinDCF}                   & \textbf{EER(\%)}                  & \textbf{MinDCF}                   & \textbf{EER(\%)} & \textbf{MinDCF} & \textbf{EER(\%)} & \textbf{MinDCF} \\ \midrule
      \textbf{Baseline}                                                                                        & ---                    & 17.29                             & 0.821                             & 6.98                              & 0.487                             & 2.99             & 0.307           & 2.30             & 0.234           \\ \midrule
      \multirow{8}{*}{\begin{tabular}[l]{@{}l@{}}\textbf{External SE}\\ \textbf{Diffusion Model}\end{tabular}} & $ \mathrm{DM}_{0.5} $  & 28.43                             & 0.887                             & 21.78                             & 0.853                             & 20.02            & 0.836           & 18.92            & 0.833           \\
                                                                                                               & $ \mathrm{DM}_{0.5+} $ & 12.43                             & 0.735                             & 5.71                              & 0.446                             & 2.89             & 0.284           & 2.24             & 0.226           \\
                                                                                                               & $ \mathrm{DM}_{1.0} $  & 22.05                             & 0.859                             & 13.19                             & 0.755                             & 12.33            & 0.730           & 12.22            & 0.730           \\
                                                                                                               & $ \mathrm{DM}_{1.0+} $ & 11.80                             & 0.712                             & 5.03                              & 0.419                             & 2.81             & 0.283           & 2.23             & 0.225           \\
                                                                                                               & $ \mathrm{DM}_{1.5} $  & 20.37                             & 0.841                             & 11.66                             & 0.709                             & 11.13            & 0.707           & 10.97            & 0.701           \\
                                                                                                               & $ \mathrm{DM}_{1.5+} $ & 10.22                             & 0.686                             & 4.93                              & \textbf{0.415}                    & \textbf{2.76}    & 0.279           & \textbf{2.21}    & \textbf{0.218}  \\
                                                                                                               & $ \mathrm{DM}_{2.0} $  & 20.11                             & 0.837                             & 11.62                             & 0.707                             & 11.07            & 0.706           & 11.01            & 0.705           \\
                                                                                                               & $ \mathrm{DM}_{2.0+} $ & \textbf{10.11}                    & \textbf{0.681}                    & \textbf{4.92}                     & 0.416                             & 2.77             & \textbf{0.277}  & 2.25             & 0.224           \\ \midrule
      \multirow{8}{*}{\begin{tabular}[l]{@{}l@{}}\textbf{Built-in SE}\\ \textbf{Diffusion Model}\end{tabular}} & $ \mathrm{DM}_{0.5} $  & 25.03                             & 0.870                             & 20.48                             & 0.843                             & 19.09            & 0.839           & 18.89            & 0.830           \\
                                                                                                               & $ \mathrm{DM}_{0.5+} $ & 10.63                             & 0.636                             & 4.95                              & 0.416                             & 2.74             & 0.273           & 2.20             & 0.225           \\
                                                                                                               & $ \mathrm{DM}_{1.0} $  & 21.95                             & 0.854                             & 12.07                             & 0.724                             & 11.88            & 0.714           & 12.16            & 0.724           \\
                                                                                                               & $ \mathrm{DM}_{1.0+} $ & 9.62                              & 0.621                             & 4.59                              & 0.392                             & 2.68             & 0.277           & 2.18             & 0.224           \\
                                                                                                               & $ \mathrm{DM}_{1.5} $  & 20.17                             & 0.837                             & 11.14                             & 0.712                             & 10.91            & 0.695           & 10.89            & 0.696           \\
                                                                                                               & $ \mathrm{DM}_{1.5+} $ & 9.39                              & 0.618                             & \textbf{4.49}                     & \textbf{0.386}                    & \textbf{2.68}    & \textbf{0.271}  & 2.18             & \textbf{0.219}  \\
                                                                                                               & $ \mathrm{DM}_{2.0} $  & 20.04                             & 0.834                             & 11.01                             & 0.707                             & 10.88            & 0.691           & 10.88            & 0.697           \\
                                                                                                               & $ \mathrm{DM}_{2.0+} $ & \textbf{9.32}                     & \textbf{0.614}                    & 4.50                              & 0.389                             & 2.69             & 0.274           & \textbf{2.16}    & 0.220           \\ \bottomrule
    \end{tabular}
  \end{adjustbox}
  \caption{Overall comparison of SV performance with SE guidance diffusion model under different short utterance scenarios.}
  \label{tab:performance}
\end{table*}

\subsection{Metrics}
\label{sec:metric}

The performance of the SV model is measured by the equal error rate (EER) and the minimum detection cost function (MinDCF) with $ \text{P}_{\text{target}}=10^{-2} $, $ \text{C}_{\text{fa}}=1 $, $ \text{C}_{\text{fr}}=1 $ \cite{desplanques2020ecapa}.

\subsection{Implementation Details}
\label{sec:implement}

The experiments are performed based on the pre-trained ECAPA-TDNN \cite{desplanques2020ecapa}.
It is used as the baseline model and the SE extractor.
Prior to the experiment, we used voice activity detection (VAD) to filter out invalid frames from the audio data.
Instead of using raw speech signals as input to the model, we utilized log Mel-filterbanks (FBanks) as the data to be synthesized.
The improvement of VoiceExtender on the SV is then examined when the audio duration is 0.5s, 1s, 1.5s, and 2s.
The gradient scale $ s $ in the external SE diffusion model and the built-in SE diffusion model is 2.0 and 3.0 respectively.
Since the experiment is tested under different short utterance durations, to achieve this, we randomly intercepted audio clips of the corresponding duration from the enroll or test audio.

It is worth noting that the SV model required for the external SE diffusion model is quite different.
Since in the reverse process, the generated samples have varying degrees of Gaussian noise, the pre-trained SV model should be robust to Gaussian noise at any signal-to-noise ratio (SNR).
Therefore, it is important to ensure that this SV model is further trained with noise (including Gaussian noise) at different SNRs to make it robust to this noise, and used for the entire experiment.

\section{Results}
\label{sec:results}

\subsection{Overall Comparison and Analysis}

In our experiments, we tested the performance of the SV model on utterance scenarios of different durations (0.5s, 1.0s, 1.5s, 2.0s) using the built-in SE and the external SE of the diffusion model, respectively, as shown in Table \ref{tab:performance}.
The baseline test shows the SV performance on audio clips of duration $ t $.
There are two types of speech included in the SV test:
(1) $ \mathrm{DM}_t $ shows the case where only the synthesized utterance of duration $ t $ is used for SV;
(2) $ \mathrm{DM}_{t+} $ shows the case where a concatenation of the original audio clip and the synthesized audio clip is used for SV.

For $ \mathrm{DM}_{t} $ series: in the scenario where the short utterance duration is 0.5s, for the external SE diffusion model, when using $ \mathrm{DM}_{0.5} $, the EER increased from 17.29\% to 28.43\% and the MinDCF increased from 0.821 to 0.887; for the built-in SE diffusion model, when using $ \mathrm{DM}_{0.5} $, the EER increased from 17.29\% to 25.03\% and the MinDCF increased from 0.821 to 0.870.
In the scenario where the short utterance duration is 1.0s: for the external SE diffusion model, when using $ \mathrm{DM}_{1.0} $, the EER increased from 6.98\% to 21.78\% and MinDCF increased from 0.487 to 0.853; for the built-in SE diffusion model, when using $ \mathrm{DM}_{1.0} $, the EER increased from 6.98\% to 20.48\% and MinDCF increased from 0.487 to 0.843.
From the above data, it can be seen that the performance of the samples generated by the diffusion model is significantly degraded compared to the original audio samples in the experiment for the same audio length.
This indicates that the samples generated by the diffusion model actually contain less audio information than the original audio samples of the same length.
Therefore, to achieve better results, the samples generated by the diffusion model need to be spliced with the original given utterance and then used for SV.

\begin{table*}[t]
  \centering
  \setlength{\tabcolsep}{8.8pt}
  \renewcommand{\arraystretch}{0.95}
  \begin{tabular}{@{}lcccccccc@{}}
    \toprule
    \multirow{3}{*}{\textbf{Given utterance}}                                                                       & \multicolumn{2}{c}{0.5s} & \multicolumn{2}{c}{1.0s} & \multicolumn{2}{c}{1.5s} & \multicolumn{2}{c}{2.0s}                                       \\ \cmidrule(l){2-9}
                                                                                                                    & EER(\%)                  & MinDCF                   & EER(\%)                  & MinDCF                   & EER(\%) & MinDCF & EER(\%) & MinDCF \\ \cmidrule(l){2-9}
                                                                                                                    & 17.29                    & 0.821                    & 6.98                     & 0.487                    & 2.99    & 0.307  & 2.30    & 0.234  \\ \midrule
    \multirow{3}{*}{\begin{tabular}[l]{@{}l@{}}\textbf{Extended utterance}\\  \textbf{by duplication}\end{tabular}} & \multicolumn{2}{c}{1.0s} & \multicolumn{2}{c}{2.0s} & \multicolumn{2}{c}{3.0s} & \multicolumn{2}{c}{4.0s}                                       \\ \cmidrule(l){2-9}
                                                                                                                    & EER(\%)                  & MinDCF                   & EER(\%)                  & MinDCF                   & EER(\%) & MinDCF & EER(\%) & MinDCF \\ \cmidrule(l){2-9}
                                                                                                                    & 17.04                    & 0.816                    & 7.11                     & 0.494                    & 2.95    & 0.305  & 2.26    & 0.232  \\ \bottomrule
  \end{tabular}
  \caption{The SV performance with given short utterance and extended utterance under different short utterance scenarios.}
  \label{tab:extended_utterance}
\end{table*}

\begin{figure}[t]
  \includegraphics[width=0.48\textwidth]{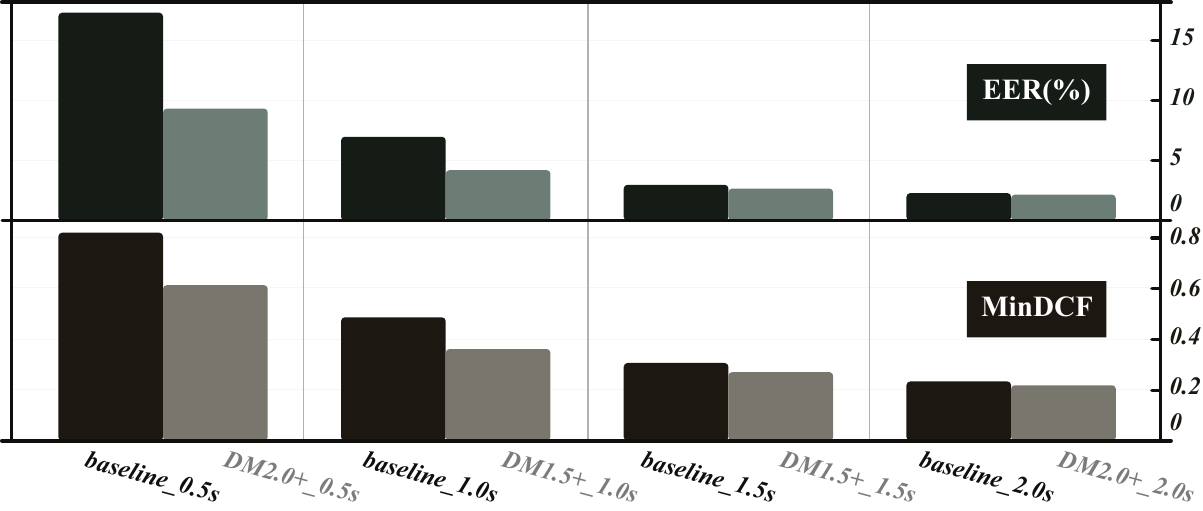}
  \centering
  \caption{Comparison of baseline\_t and the best DMt+ on EER and MinDCF when t takes values of (0.5s,1.0s,1.5s and 2.0s).}
  \label{fig:compare}
\end{figure}

For $ \mathrm{DM}_{t+} $ series:
in the scenario where the short utterance duration is 0.5s, for the external SE diffusion model, when using $ \mathrm{DM}_{2.0+} $, EER decreased from 17.29\% to 10.11\% and MinDCF dropped from 0.821 to 0.681; for the built-in SE diffusion model, when using $ \mathrm{DM}_{1.5+} $, EER dropped from 17.29\% to 9.32\% and MinDCF dropped from 0.821 to 0.614.
In the scenario where the short utterance duration is 1.0s, for the external SE diffusion model, when using $ \mathrm{DM}_{2.0+} $, the EER dropped from 6.98\% to 4.92\% and MinDCF dropped from 0.487 to 0.416; for the built-in SE diffusion model, when using $ \mathrm{DM}_{1.5+} $, EER dropped from 6.98\% to 4.49\% and MinDCF dropped from 0.487 to 0.362.
Similarly, we can see that both SE-guided diffusion models improve the SV performance to different degrees in the 1.5s and 2.0s scenarios.
For easier comparison, we have extracted a detailed comparison of the baseline and optimal diffusion models on the EER and MinDCF, see Figure \ref{fig:compare}.
The results show that after using VoiceExtender, the performance of SV is significantly improved in different utterance duration scenarios.

In summary, the SV model achieves best-case performance gains of (41.5\% 29.5\% 7.6\% 3.9\%; 46.1\% 35.7\% 10.4\% 5.7\%) in EER and (17.1\% 14.6\% 9.1\% 4.3\%; 25.2\% 20.7\% 11.7\% 7.3\%) in MinDCF when using external and built-in SE diffusion models, respectively.
The built-in SE diffusion model performs better than the external SE diffusion model in improving the speaker SV performance.
This may be due to the fact that the external SE diffusion model requires better SV models that are robust to noise.
Furthermore, it can be noticed that the enhancement of the diffusion model on the SV performance becomes more pronounced as the speech duration becomes shorter, which means that VoiceExtender is more effective for extremely short utterances.

\subsection{Duplicated Audio Baseline Comparison}

To investigate the effect of audio duration alone on SV performance, without introducing additional information, experiments are conducted in which the original audio is duplicated to double its length across the 4 duration scenarios (see Table \ref{tab:extended_utterance}).
Extending duration by simple duplication had minimal effect, suggesting that diffusion models generate samples with substantial novel speech information that improves SV.
The key reason underlying this finding is that well-trained diffusion models synthesize samples containing meaningful additional speech content that enhances SV capabilities, rather than simply extending duration.

\begin{figure}[t]
  \includegraphics[width=0.48\textwidth]{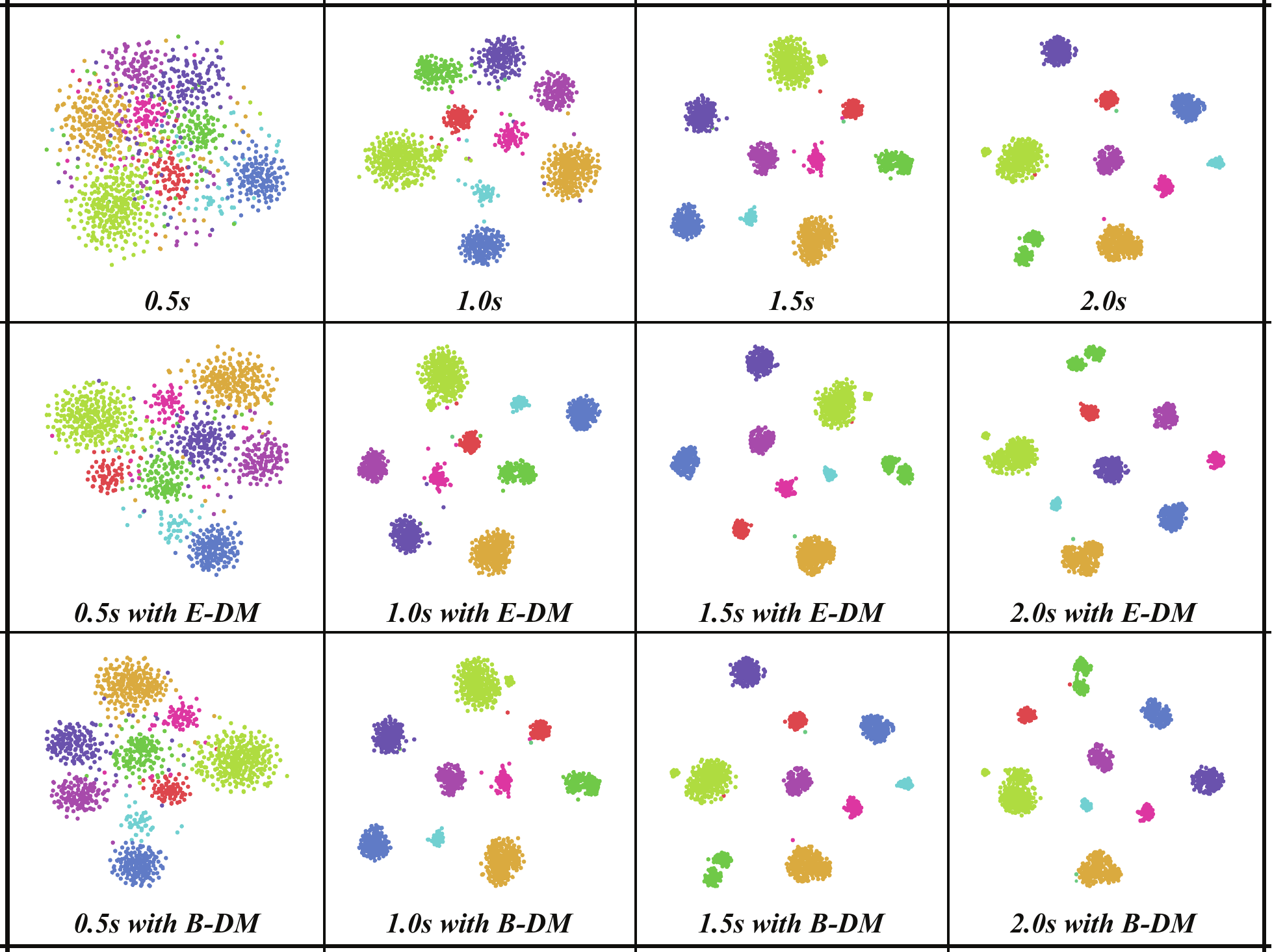}
  \centering
  \caption{Visualization of SEs for top ten speakers in the test dataset using t-SNE: B-DM means built-in SE diffusion model; E-DM means external SE diffusion model.}
  \label{fig:tsne}
\end{figure}

\subsection{Generated Samples Visualization}

To further exhibit VoiceExtender's enhancement of SV performance, t-SNE visualizations of the top 10 speakers' SEs on the test set were generated (see Figure \ref{fig:tsne}).
The results demonstrate that VoiceExtender can increase inter-speaker SE separation and intra-speaker SE aggregation.
This effect becomes more pronounced as utterance duration decreases.

\section{Conclusions}

In this paper, we propose VoiceExtender, which is an effective and promising framework for improving the performance of speaker verification (SV) when dealing with short utterances.
By employing a diffusion model with built-in or external SE guidance, VoiceExtender is able to generate additional speech features from a given short utterance, thereby extending its duration and improving the performance of the speaker verification (SV) system.
Experimental results on the Voxceleb dataset show that the VoiceExtender architecture outperforms the baseline model, with relative improvements of 46.1\%, 35.7\%, 10.4\%, and 5.7\% in EER and 25.2\%, 20.7\%, 11.7\%, and 7.3\% in MinDCF under 0.5, 1.0, 1.5, and 2.0 seconds short-utterance conditions.

\section{Acknowledgement}

Supported by the Key Research and Development Program of Guangdong Province (grant No. 2021B0101400003) and Corresponding author is Jianzong Wang (jzwang@188.com).


\bibliographystyle{IEEEbib}
\bibliography{refs}

\end{document}